\documentclass[conference]{IEEEtran}
\IEEEoverridecommandlockouts
\usepackage{cite}
\usepackage{amsmath,amssymb,amsfonts}
\usepackage{graphicx}
\usepackage{textcomp}
\usepackage{xcolor}
\usepackage{siunitx}
\usepackage{algorithm}   
\usepackage{algorithmic}
\usepackage{multirow}
\usepackage{multicol}
\usepackage{graphicx}
\usepackage{booktabs}
\usepackage{colortbl}
\usepackage{xcolor}

\def\BibTeX{{\rm B\kern-.05em{\sc i\kern-.025em b}\kern-.08em
    T\kern-.1667em\lower.7ex\hbox{E}\kern-.125emX}}
\begin{document}

\title{Beyond Autoregression: Diffusion Language Model is All Analog In-Memory Computing Need}

\author{
\IEEEauthorblockN{Yuanuo Feng$^{\dagger 1}$, Yubing Ye$^{\dagger 2}$, Yizhe Chen$^{1}$, Wenshuai Yao$^{3}$, Ngai Wong$^{2}$, Wenyong Zhou$^{*2}$, Wang Kang$^{*1}$,}
\IEEEauthorblockA{
$^1$School of Integrated Circuit Science and Engineering, Beihang University, Beijing, China\\
$^2$Faculty of Engineering, The University of Hong Kong, Hong Kong\\
$^3$School of Integrated Circuits, Peking University, Beijing, China \quad
\textsuperscript{\dag}: Equal contribution. \textsuperscript{*}: Corresponding authors.
}
}
\maketitle

\begin{abstract}
Analog in-memory computing (AIMC) enables energy-efficient language-model inference, but hardware non-idealities perturb matrix--vector multiplications and degrade generation quality. Autoregressive (AR) models are especially vulnerable: each token is committed to the causal context at birth, so perturbation-induced errors cascade. We study diffusion language models (DLMs), which iteratively refine a masked sequence, as an alternative paradigm. Across scale-matched model pairs and open checkpoints, DLMs degrade more gracefully than AR counterparts under persistent weight perturbations. Controlled analyses rule out cross-round noise averaging and support delayed token commitment and progressive context resolution as the mechanism. A larger denoising-step budget further shrinks the clean--noisy gap, a DLM-specific latency--robustness knob. We perform large-scale simulation tests across diverse datasets and models using a noise model calibrated from real silicon measurements. Furthermore, on a 16-chip 28nm eFlash AIMC cluster, hybrid on-board measurements show DLMs tolerate $1.4$--$3\times$ larger weight noise, making diffusion-based generation a promising paradigm for noise-resilient AIMC inference.

\end{abstract}


\section{Introduction}

Large language models (LLMs)~\cite{brown2020language} are increasingly deployed in applications that require low-latency~\cite{DBLP:conf/mlsys/PopeDCDBHXAD23} and energy-efficient generation~\cite{DBLP:conf/acl/FernandezNTBLS25}. Analog in-memory computing (AIMC) is a promising substrate for this workload, performing matrix--vector multiplications (MVMs) directly in memory and reducing data movement~\cite{DBLP:journals/ncs/LerouxMSFSSN25}. However, AIMC introduces device- and circuit-level non-idealities, including programming variability, conductance drift, read noise, and peripheral-circuit imperfections~\cite{rasch2023hardware}, which perturb the effective weights and thus the output of each MVM~\cite{joshi2020accurate}. Although such perturbations may be small at the operation level, they can accumulate and amplify throughout language generation~\cite{DBLP:conf/date/EldebikyZBLS23, DBLP:conf/nips/BuchelCACFTMGRS25, DBLP:journals/ncs/LerouxMSFSSN25}.

This amplification is especially problematic for autoregressive (AR) generation. AR models generate tokens from left to right, and each generated token is immediately incorporated into the causal context used to predict all subsequent tokens. Consequently, a perturbation-induced error at an early position can alter the context observed by later positions and trigger an error cascade over the rest of the sequence~\cite{DBLP:conf/acl/AroraABC22,  DBLP:journals/corr/RanzatoCAZ15, DBLP:conf/icml/ZhangPMLS24}. AR robustness on AIMC therefore hinges not only on per-operation or per-token error, but on whether a local error is irreversibly committed to the generation trajectory.

\begin{figure}[!t]
\centering
\includegraphics[width=1.0\columnwidth]{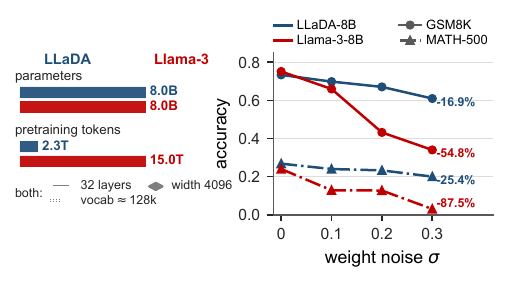}
\caption{Comparison of two 8B-scale language models under  weight
perturbations. LLaDA-8B and Llama-3-8B share depth (32 layers), width
(4096), and vocabulary size (${\approx}$128k).}
\label{fig:llada_vs_llama}
\vspace{-0.5cm}
\end{figure}

Diffusion language models (DLMs) offer a different generation process~\cite{DBLP:conf/nips/NieZYZOHZLWL25}: rather than committing tokens strictly from left to right, they iteratively refine a partially masked sequence~\cite{DBLP:conf/nips/AustinJHTB21, DBLP:conf/nips/SahooASGMCRK24} and can defer uncertain decisions while resolving more reliable positions~\cite{DBLP:conf/icml/KimSKKC25}. Fig.~\ref{fig:llada_vs_llama} shows that this difference has direct hardware consequences: under  weight perturbations, although Llama-3-8B is pretrained on $6.5\times$ more tokens, LLaDA-8B degrades much more gracefully on both benchmarks and overtakes Llama-3 on GSM8K at moderate noise levels. This robustness is therefore a distinct generation characteristic rather than a consequence of training scale.

\begin{figure*}[!t]
\centering
\includegraphics[width=0.95\textwidth]{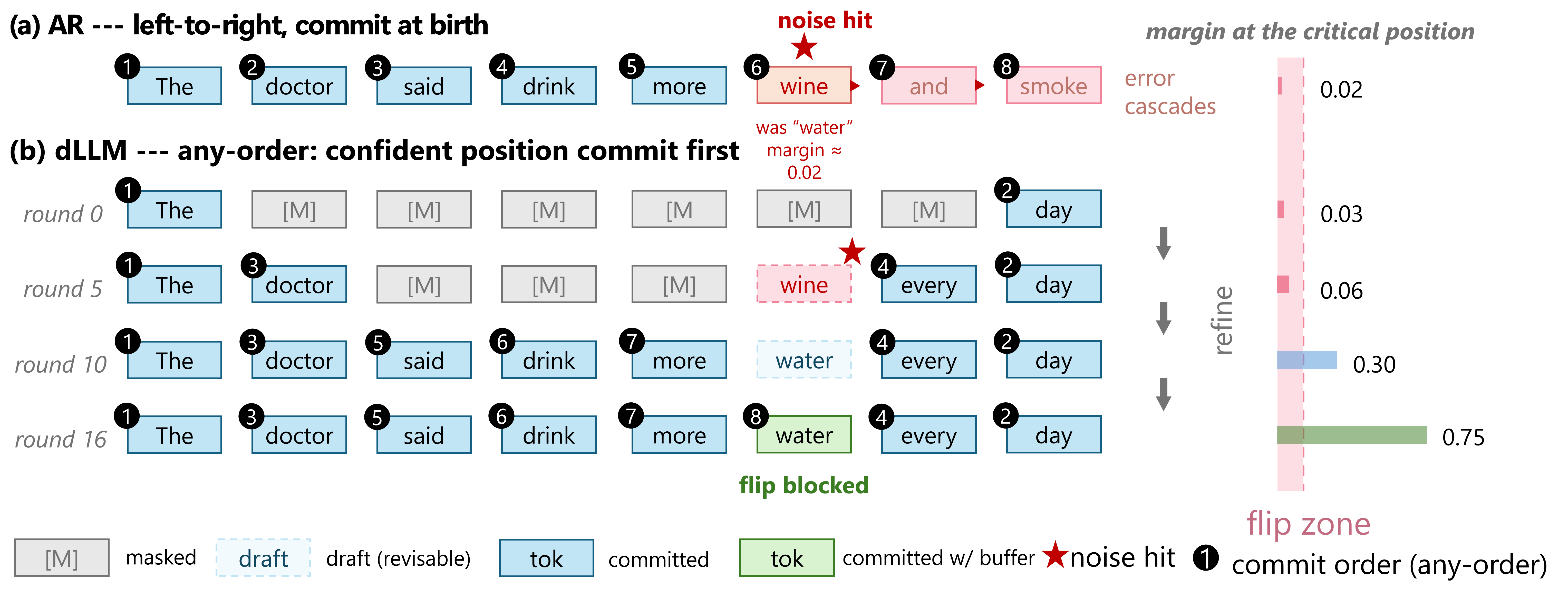}
\caption{Commitment timing under weight noise: the reason behind the
robust behavior in Fig.~\ref{fig:llada_vs_llama}, as identified by the
controlled analyses in Sec.~\ref{sec:generation_protocols} and
Sec.~\ref{sec:context_resolution}. An AR decoder (top row) commits each
token immediately at birth, so an early noise-induced flip
(\textit{water}$\rightarrow$\textit{wine}) enters the causal context and
cascades through subsequent predictions. A masked DLM (bottom rows)
commits confident positions first and keeps uncertain positions masked or
revisable; as more context is resolved, the prediction margin at the
critical position recovers and the flip is blocked.}
\label{fig:reason}
\vspace{-0.5cm}
\end{figure*}
Fig.~\ref{fig:reason} illustrates a candidate mechanism for this behavior. In AR decoding, each token is committed at birth: a noise-induced flip from \textit{water} to \textit{wine} immediately enters the causal context and cascades. A masked DLM instead commits confident positions first and keeps uncertain positions revisable, so a perturbation that initially favors an incorrect token can be reconsidered as more context is resolved, and the recovered prediction margin blocks the flip. It is unclear a priori, however, whether the observed robustness comes merely from iterative denoising and inter-round noise averaging, or from these commitment and error-propagation dynamics. Answering this question is critical for hardware-algorithm co-design: if delayed commitment is the source, DLMs can tolerate larger persistent weight perturbations and relax hardware fidelity demands. The analyses in Sec.~\ref{sec:robustness_analysis} test these alternatives and support the commitment-based explanation.

In this work, we study DLM robustness under AIMC-induced weight perturbations, progressing from controlled analysis to calibrated simulation and on-board measurement. We first show that the tested DLM checkpoints degrade more gracefully than the tested AR checkpoints across scale-matched models and open checkpoints on PIQA~\cite{DBLP:conf/aaai/BiskZLGC20} and GSM8K~\cite{DBLP:journals/corr/abs-2110-14168}. Controlled commitment interventions and context analyses then identify delayed commitment, rather than cross-round noise averaging, as a supported mechanism for this behavior. We finally quantify the resulting hardware advantage through a chip-calibrated perturbation model and hybrid measurements on a 16-chip AIMC cluster.
Our contributions are as follows:
\begin{itemize}
    \item We show that DLMs degrade more gracefully than AR models under AIMC-induced weight perturbations across scale-matched models and downstream benchmarks.

    \item We show that this robustness is not explained by cross-round noise averaging, and identify delayed token commitment and progressive context resolution as supported mechanisms.

    \item We demonstrate the hardware advantage on a 16-chip 28nm eFlash AIMC cluster: under a perturbation model calibrated against on-chip MVM errors, an 8B-scale sweep on eight benchmarks and hybrid on-board measurements (transformer linear MVMs on-chip, the rest digital) show that DLMs tolerate $1.4$--$3\times$ larger weight noise, with additional denoising steps trading measured latency and energy for a smaller clean--noisy gap, a DLM-specific latency--robustness knob.
\end{itemize}

\section{Related Work}

\noindent\textbf{Analog in-memory computing.}
AIMC accelerates neural-network inference by performing matrix--vector multiplications within memory arrays, reducing data movement and improving energy efficiency~\cite{shafiee2016isaac, khaddam2022hermes}. However, device variability, conductance drift, finite programming precision, and peripheral-circuit non-idealities perturb the effective weights and degrade inference accuracy~\cite{joshi2020accurate, rasch2023hardware}. Prior work has addressed these effects through hardware-aware training, calibration, compensation, and noise-robust architectures~\cite{rasch2023hardware, DBLP:conf/date/EldebikyZBLS23, DBLP:journals/tcad/WangCLLFMWZWZPTK26}. These studies primarily focus on discriminative workloads or token-level accuracy, whereas the effect of AIMC perturbations on the sequential error dynamics of modern language generation remains less understood.

\noindent\textbf{Robustness in autoregressive generation.}
Autoregressive language models generate tokens sequentially, conditioning each prediction on previously generated tokens. This decoding process can amplify early mistakes through exposure bias and error propagation~\cite{DBLP:conf/nips/BengioVJS15, DBLP:journals/corr/RanzatoCAZ15, DBLP:conf/acl/AroraABC22}. Existing approaches mitigate such errors through improved training objectives, decoding strategies, or post-generation refinement. In contrast, we study errors introduced by imperfect inference hardware: a perturbation-induced token flip can be irrevocably inserted into the causal context, causing subsequent predictions to condition on a corrupted sequence.

\noindent\textbf{Diffusion language models.}
Diffusion language models generate text through iterative denoising rather than fixed left-to-right decoding. Discrete and masked diffusion methods have demonstrated increasingly competitive language-modeling and generation quality, while enabling parallel or confidence-driven token updates~\cite{DBLP:conf/nips/AustinJHTB21, DBLP:conf/icml/LouME24, DBLP:conf/nips/SahooASGMCRK24, DBLP:conf/nips/NieZYZOHZLWL25, ghazvininejad2019mask, DBLP:conf/icml/KimSKKC25}. Recent work has also explored improved sampling, self-correction, and scaling behavior for DLMs~\cite{DBLP:conf/nips/WangSSK25, DBLP:conf/iclr/NieZDPLZLL25}. These efforts focus on generation quality, sampling efficiency, or controllability under ideal computation. Our work instead studies DLMs under AIMC-induced weight perturbations. We show that their advantage is not explained by cross-round noise averaging, but by delayed token commitment: uncertain tokens remain revisable until a more reliable context is available, thereby suppressing hardware-induced error cascades.


\section{Robustness Evaluation and Analysis}
\label{sec:robustness_analysis}

We evaluate the robustness of diffusion language models (DLMs) under persistent
weight perturbations and compare their behavior with autoregressive (AR) models.

\subsection{ weight Perturbation and Robustness Evaluation}
\label{sec:static_noise_eval}

We consider inference-time weight perturbations. Let $\theta$
denote the clean model parameters and let $\Delta\theta$ denote a perturbation
sampled at noise level $\sigma$. The perturbed parameters are
\begin{equation}
    \tilde{\theta} = \theta + \Delta\theta.
    \label{eq:static_perturbation}
\end{equation}
For every mapped parameter, the calibrated perturbation follows
\begin{equation}
 \Delta\theta_{ij}\sim\mathcal{N}\!\left(
 0,(k_{\mathrm{std}}|\theta_{ij}|+b_{\mathrm{std}})^2\right).
 \label{eq:noise_distribution}
\end{equation}
The GPU experiments use the weight-dominant limit
$b_{\mathrm{std}}\!\rightarrow0$, which is equivalent to
$\tilde{\theta}=\theta\odot(1+\sigma Z)$ with independent
$Z_{ij}\sim\mathcal{N}(0,1)$. All mapped linear-layer weights and biases are
perturbed; embeddings and the language-model head remain digital and are
excluded. This notation reconciles the additive physical model with the
effective multiplicative perturbation used in simulation.
For each benchmark evaluation job, one error map is sampled per noise seed and reused for every sample and every round of that job. It therefore models one programmed hardware instance; sampler randomness is independent of the error map. The perturbation remains fixed across all AR decoding steps or DLM denoising rounds:
\begin{equation}
    \tilde{\theta}^{(r)} = \tilde{\theta},
    \qquad \forall r,
    \label{eq:shared_noise}
\end{equation}
where $r$ indexes an AR decoding step or a DLM denoising round. This setting
models persistent programming and device mismatch, isolating the effect of a
fixed programmed error map on the full generation trajectory.

GSM8K uses flexible numerical extraction; LLaDA and Llama use 5-shot and
8-shot prompts, respectively, following their released evaluation protocols;
the matched-model comparison below complements this open-checkpoint study
with fully aligned training and evaluation settings.
The primary noise sweep covers the full GSM8K test set
(1{,}319 questions) with seven noise levels and three noise seeds; PIQA
studies use the full validation split (1{,}838 examples).
The workspace includes sample-level outputs, three to five noise seeds where
available, four training seeds for the matched small models, and aggregation
scripts in \texttt{scripts/} and \texttt{training\_causality/}, providing
seed-level resolution for every reported trend.

For a performance metric $A$, we define the noise-induced degradation of a model
$M$ at perturbation level $\sigma$ as
\begin{equation}
    D_M(\sigma) = A_M(0) - A_M(\sigma),
    \label{eq:performance_degradation}
\end{equation}
where $A_M(0)$ and $A_M(\sigma)$ are the clean and perturbed performance,
respectively. Smaller values of $D_M(\sigma)$ indicate greater robustness to
 weight perturbations. When evaluating validation loss $\mathcal{L}$, we
instead report the relative loss increase
\begin{equation}
    R_M(\sigma)
    =
    \frac{\mathcal{L}_M(\sigma)-\mathcal{L}_M(0)}
    {\mathcal{L}_M(0)}.
    \label{eq:relative_loss_increase}
\end{equation}

\begin{figure}[!t]
\centering
\includegraphics[width=1.0\columnwidth]{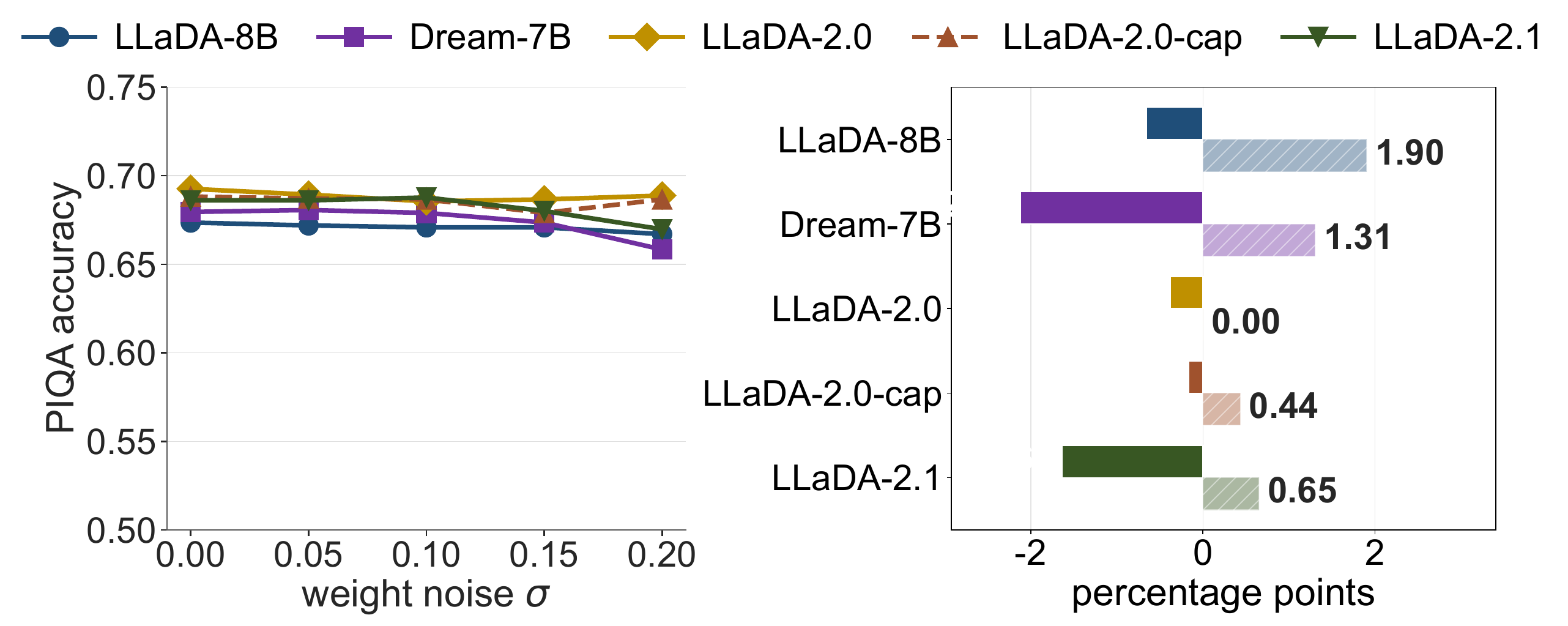}
\caption{Robustness of open DLM checkpoints on PIQA under  weight
perturbations. \textbf{Left:} accuracy as a function of perturbation strength
$\sigma$. \textbf{Right:} accuracy differences at $\sigma=0.2$, decomposed into
differences already present under clean weights and the additional degradation
induced by perturbations. }
\label{fig:piqa_noise}
\end{figure}

We first evaluate several open DLM checkpoints on PIQA. Fig.~\ref{fig:piqa_noise}
shows accuracy across perturbation strengths and separates noisy-model
differences into clean-model gaps and perturbation-induced degradation at
$\sigma=0.2$.

The decomposition in Fig.~\ref{fig:piqa_noise} is important because a
performance gap between noisy models does not by itself establish a robustness
difference: part of the gap may already be present in the clean setting.
Accordingly, we distinguish absolute perturbed performance from the degradation
introduced by noise.

\begin{figure}[!t]
\centering
\includegraphics[width=0.95\columnwidth]{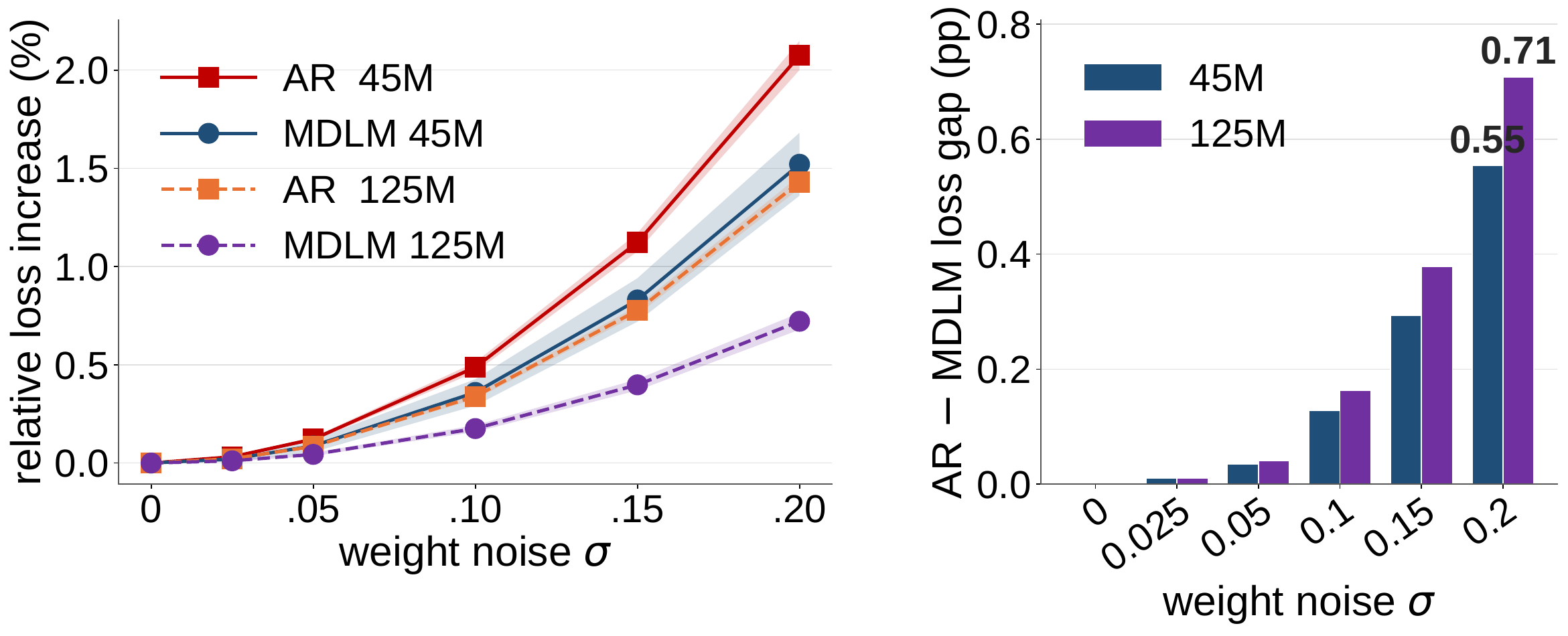}
\caption{Controlled comparison between matched AR and MDLM models under weight perturbations. \textbf{Left:} relative validation-loss increase as a
function of perturbation strength $\sigma$. \textbf{Right:} the excess relative
loss increase of the AR model over the MDLM model.}
\label{fig:matched_ar_mdlm}
\end{figure}

We next perform a controlled comparison between matched AR and MDLM models.
Because the two models are matched in scale and evaluated under aligned training
and perturbation settings, this comparison isolates robustness differences that
cannot be attributed solely to model size or clean-model quality.
Fig.~\ref{fig:matched_ar_mdlm} reports the relative validation-loss increase
defined in Eq.~\eqref{eq:relative_loss_increase}.

The MDLM exhibits a smaller noise-induced relative loss increase than its AR
counterpart. Together with the open-checkpoint evaluation, this result indicates
that the observed robustness is not merely a consequence of a particular model
checkpoint or an advantage in clean performance.

\subsection{Controlled Generation Protocols}
\label{sec:generation_protocols}

A possible alternative explanation for DLM robustness is that repeated
denoising rounds average out independent parameter perturbations. This mechanism
would not apply to the default noise setting in
Eq.~\eqref{eq:shared_noise}, where the same perturbed model is used throughout
the complete trajectory. Nevertheless, we explicitly test this explanation by
comparing shared perturbations with round-wise resampled perturbations.

The shared-noise condition follows Eq.~\eqref{eq:shared_noise}; the
resampled-noise condition draws an independent perturbation at every round:
\begin{equation}
    \tilde{\theta}^{(r)}
    =
    \theta + \Delta\theta^{(r)},
    \qquad
    \Delta\theta^{(r)} \sim \mathcal{P}_{\sigma},
    \label{eq:resampled_perturbation}
\end{equation}
where $\mathcal{P}_{\sigma}$ denotes the perturbation distribution at noise
level $\sigma$. If robustness were primarily caused by averaging independent
noise across rounds, the resampled-noise condition should yield a systematic
advantage over shared noise.

\begin{figure}[!t]
\centering
\includegraphics[width=1.0\columnwidth]{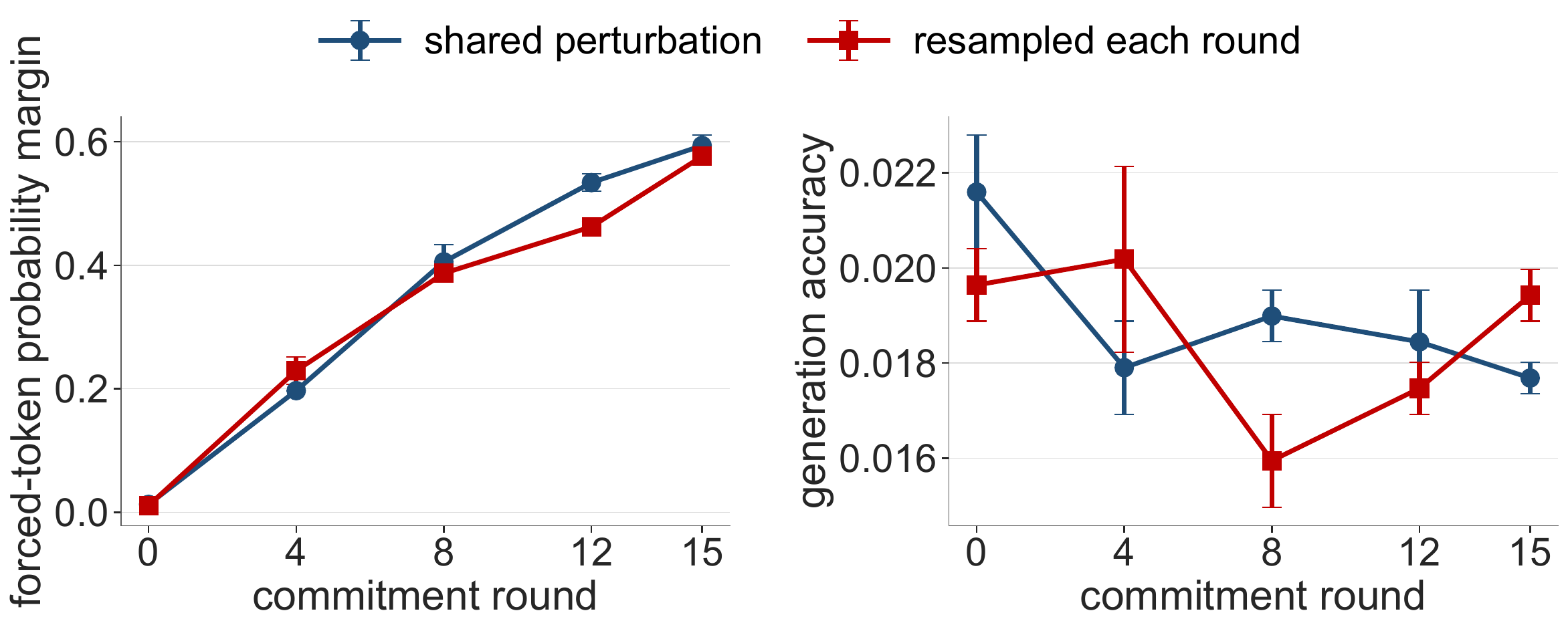}
\caption{Effect of perturbation scheduling during DLM generation. In the
shared-noise condition, one  weight perturbation is retained across all
denoising rounds; in the resampled-noise condition, perturbations are
independently redrawn at each round. \textbf{Left:} forced-token margin
across commitment rounds. \textbf{Right:} generation accuracy across
commitment rounds.}
\label{fig:independent_noise}
\end{figure}

Fig.~\ref{fig:independent_noise} compares the two schedules using both the
forced-token margin and generation accuracy across commitment rounds.

Round-wise resampling does not provide a systematic advantage over the shared-noise condition. This rules out independent round-wise averaging as a sufficient explanation, but does not by itself identify a unique mechanism. The commitment interventions below directly test whether changing the commitment time changes post-intervention divergence.

\begin{figure}[!t]
\centering
\includegraphics[width=1.0\columnwidth]{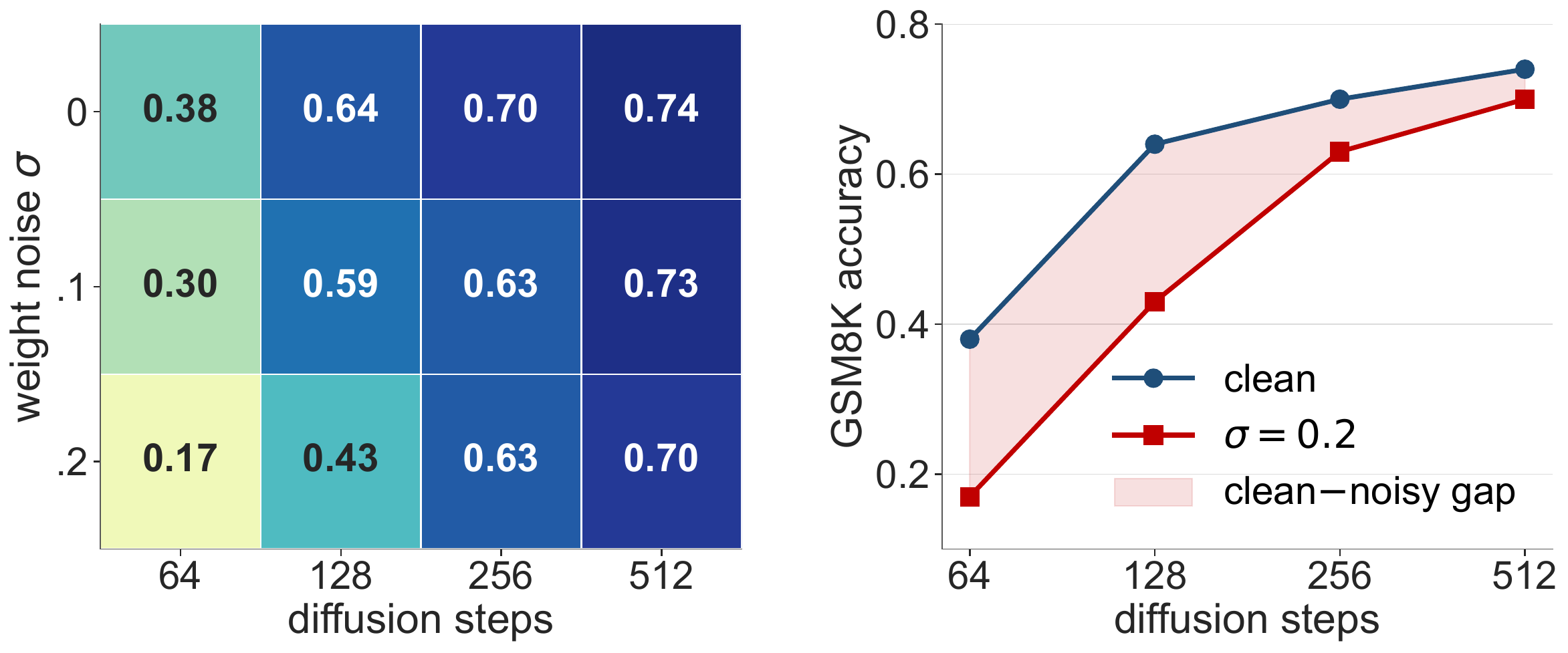}
\caption{GSM8K accuracy under different diffusion-step budgets $K$ and weight perturbation strengths $\sigma$. Increasing the denoising budget reduces
the clean--noisy performance gap.}
\label{fig:step_budget}
\end{figure}

We further vary the diffusion-step budget $K$ to determine whether additional
denoising rounds specifically improve robustness. For each step budget and noise
level, we measure the clean--noisy performance gap
\begin{equation}
    D(K,\sigma) = A(K,0) - A(K,\sigma).
    \label{eq:step_budget_gap}
\end{equation}
This quantity separates a generic improvement in clean generation from a
reduction in sensitivity to noisy weights. Fig.~\ref{fig:step_budget} reports
GSM8K accuracy across diffusion-step budgets and perturbation levels.

As the diffusion-step budget increases, the gap between clean and perturbed
performance decreases. Therefore, additional denoising rounds do more than
improve clean-generation quality: they create further opportunities to correct
or defer uncertain predictions before final token decisions are made.

\subsection{Context-Resolution Analysis}
\label{sec:context_resolution}

To characterize the mechanism behind this behavior, we analyze how resolved
context affects token-level confidence under  weight noise. For a target
position $i$, let $x_i^\star$ denote the reference token and let
$\mathcal{C}_i(q)$ denote the available context when a fraction $q$ of relevant
positions has been resolved. We define the reference-token log-probability
margin as
\begin{equation}
    m_i(q,\sigma)
    =
    \log p_{\tilde{\theta}}
    \left(x_i^\star \mid \mathcal{C}_i(q)\right)
    -
    \max_{v \neq x_i^\star}
    \log p_{\tilde{\theta}}
    \left(v \mid \mathcal{C}_i(q)\right).
    \label{eq:reference_margin}
\end{equation}
A positive margin indicates that the reference token is ranked above every
alternative token. Conversely, a margin close to zero indicates a flip-prone
decision, for which a small perturbation may change the top-ranked token.

To isolate the local effect of weight noise, we compute the margin shift with
respect to clean weights:
\begin{equation}
    \Delta_{\mathrm{noise}}(q,\sigma)
    =
    m(q,\sigma)-m(q,0).
    \label{eq:noise_margin_shift}
\end{equation}
Negative values indicate that perturbations reduce confidence in the
reference token. We compare this penalty with the margin gain obtained as
additional context becomes resolved:
\begin{equation}
    G_{\mathrm{ctx}}(\sigma)
    =
    m(q_{\max},\sigma)-m(q_{\min},\sigma).
    \label{eq:context_margin_gain}
\end{equation}

\begin{figure}[!t]
\centering
\includegraphics[width=1.0\columnwidth]{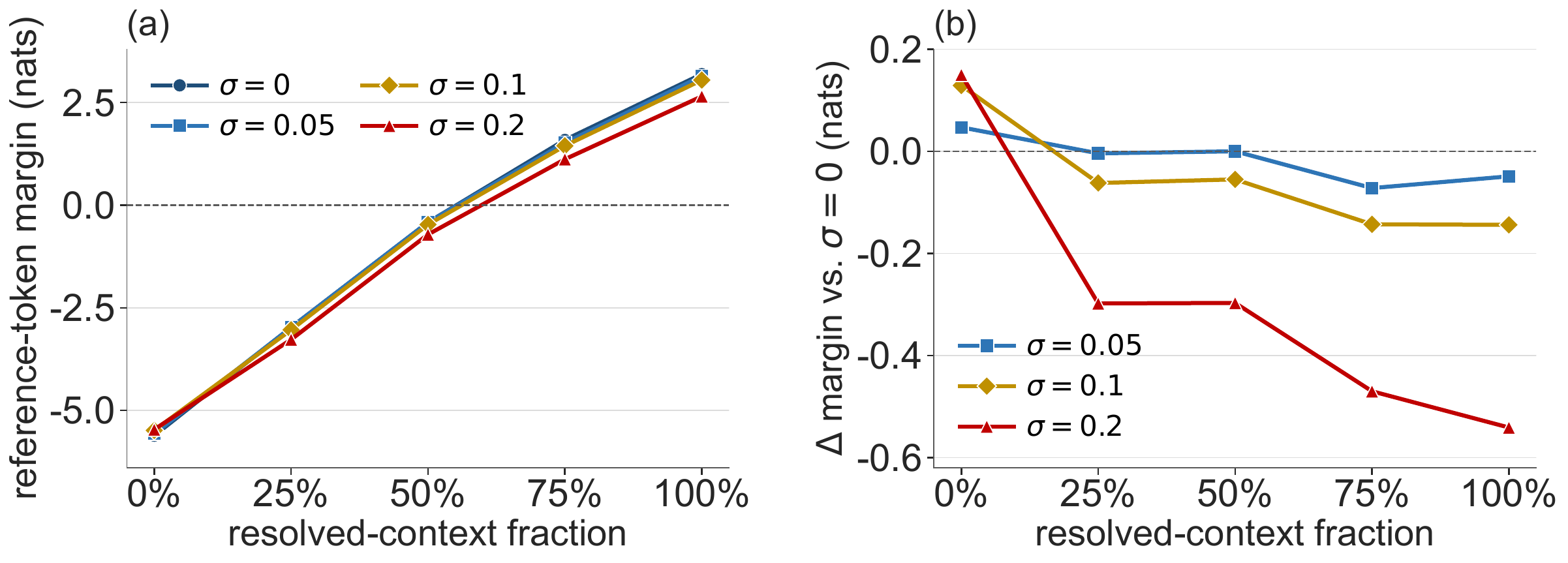}
\caption{Resolved-context analysis under  weight perturbations.
\textbf{Left:} reference-token log-probability margin as a function of the
resolved-context fraction. \textbf{Right:} the margin shift induced by
weight perturbations relative to the clean condition ($\sigma=0$).}
\label{fig:context_quality}
\end{figure}

Fig.~\ref{fig:context_quality} compares the effect of context resolution with
the direct local effect of  weight perturbations.

The reference-token margin increases monotonically as more oracle context is resolved, under both clean and perturbed weights. In contrast,  weight noise causes
a comparatively small downward margin shift, with the largest observed shift
equal to $-0.541$ nats. The confidence gain supplied by resolved context is
therefore substantially larger than the local confidence loss caused by the
perturbation.

Because this experiment reveals reference tokens from the ground-truth sequence, it tests a local margin hypothesis rather than the full natural rollout. In a complementary 45M study, natural-rollout context produced comparable or larger forced-token margins than oracle context (0.30 vs. 0.30 at fraction 8; 0.56 vs. 0.33 at fraction 12), whereas random context collapsed the margin to 0.02. Together with the commitment intervention, these results are consistent with delayed commitment and progressive context resolution reducing error propagation; they do not establish that every natural rollout has oracle-quality context.

Paired interventions hold the prompt, noise seed, and total denoising
budget fixed while varying the forced-commitment round. Delaying commitment
from round 0 to round 8 reduces post-force mismatch from 0.835 to 0.265
in the 45M model and from 0.85 to 0.30 in the 125M model. This supports a
role for commitment timing in error propagation without establishing it as
the sole source of robustness.

\section{Exploiting DLM Robustness on AIMC}
\label{sec:application}


\subsection{Hardware Platform and Cluster Deployment}
\label{sec:platform}

\begin{figure}[!t]
\centering
\includegraphics[width=1.0\columnwidth]{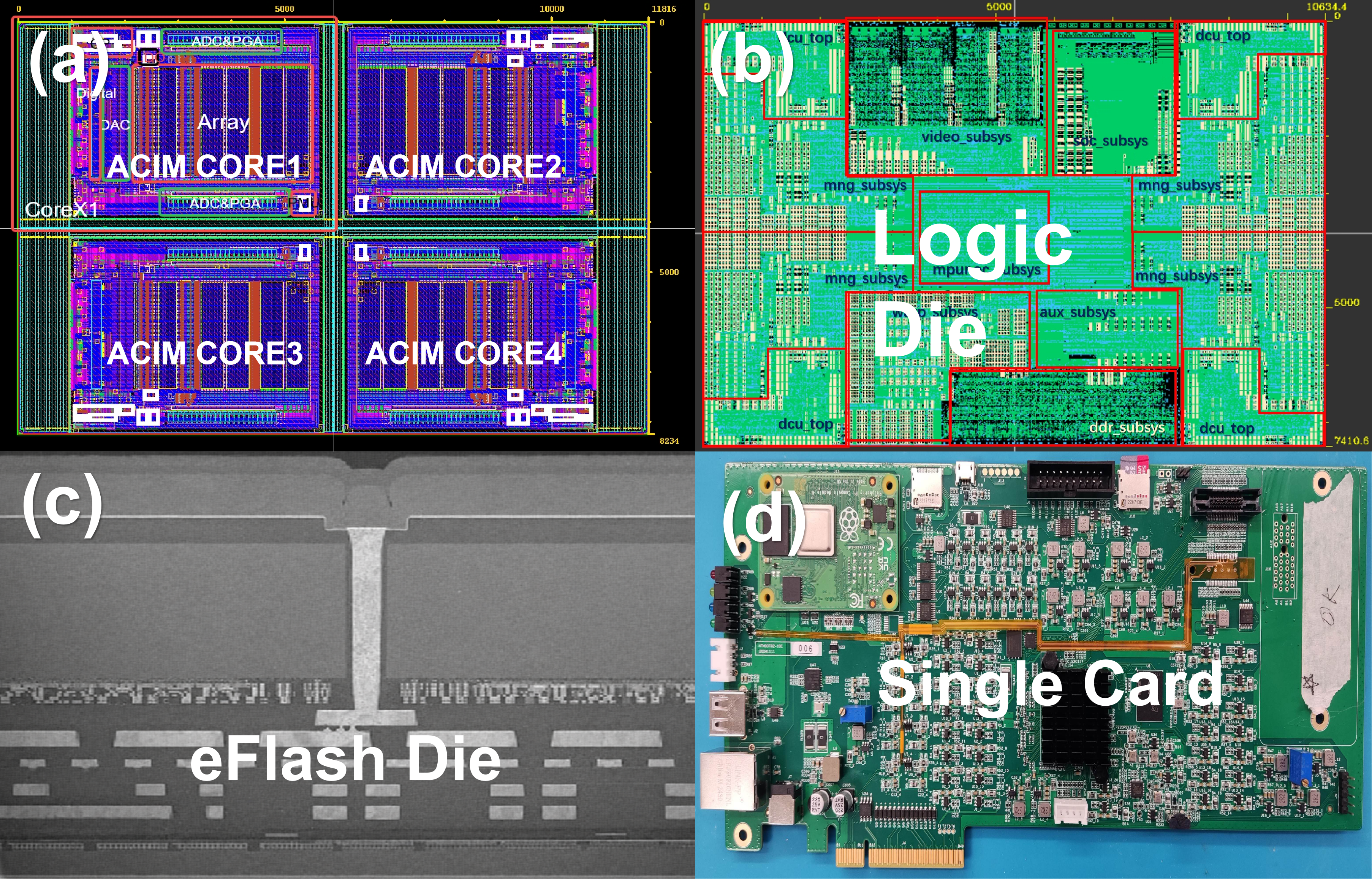}
\caption{The 28nm eFlash-based AIMC prototype. (a) AIMC die with four
cores, each integrating weight-stationary arrays with DAC and ADC/PGA
peripherals. (b) Logic die layout integrating PCIe/DDR, video,
management, and MPU/NoC subsystems. (c) TEM cross-section of the
hybrid-bonded 3D stack (12nm logic die, TSV layer, and 28nm eFlash
die). (d) Assembled single-card prototype.}
\label{fig:chipout}
\end{figure}

Our hardware platform is a 28nm eFlash-based AIMC prototype
(Fig.~\ref{fig:chipout}) with four eFlash compute cores, DAC/ADC peripherals,
and a 12nm logic die for control and PCIe/DDR, hybrid-bonded into a 3D
stack. We use the 16-chip cluster in a
hybrid measurement flow; Table~\ref{tab:onboard} defines its computation
boundary. Persistent programming and device mismatch motivate the static
perturbation in Eq.~\eqref{eq:static_perturbation}.

\begin{figure}[!t]
\centering
\includegraphics[width=1.0\columnwidth]{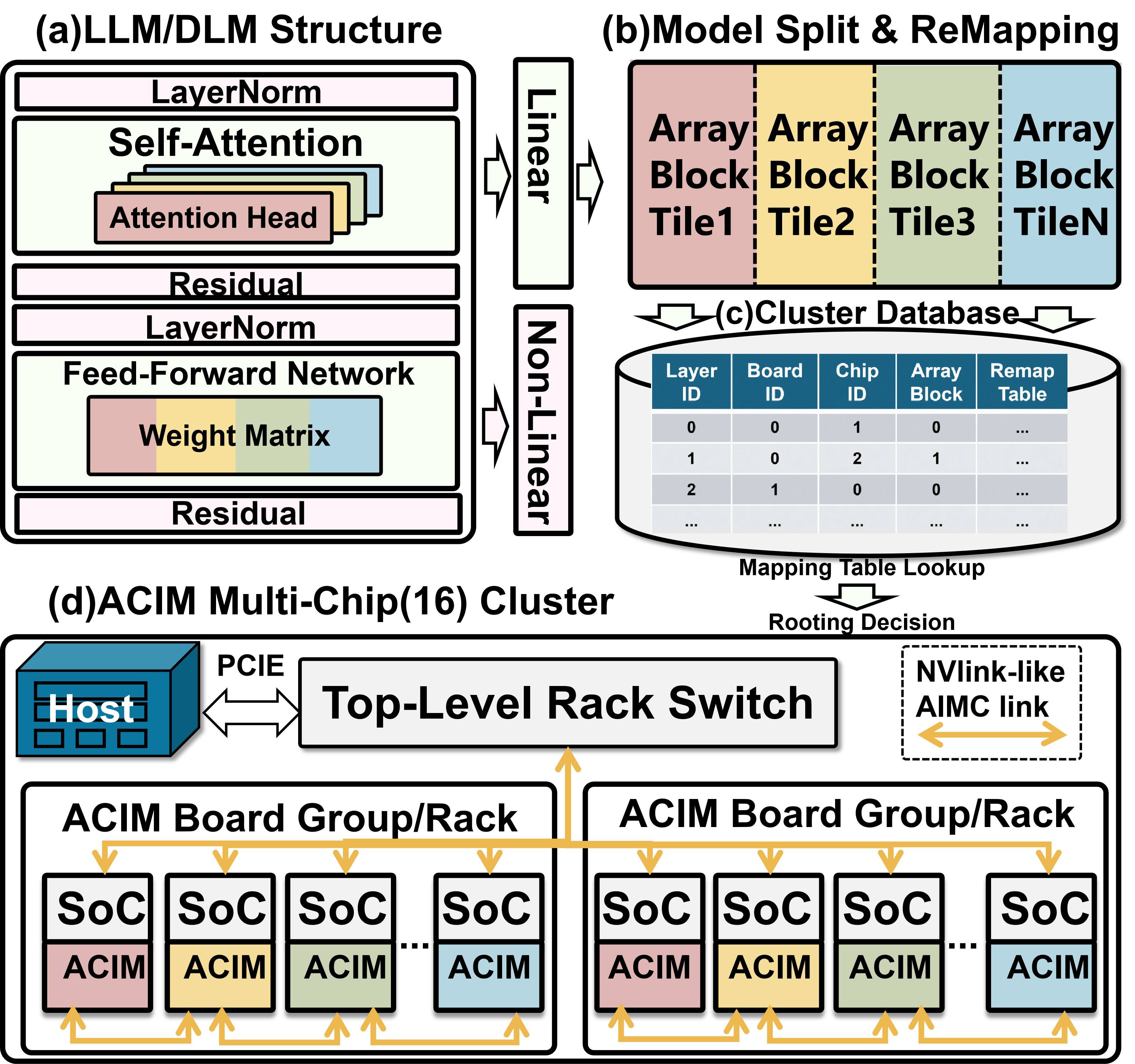}
\caption{Model deployment on the AIMC cluster. (a) Linear layers as
mapping targets. (b) Array-block tiling and remapping around defective
columns. (c) Cluster database for layer-to-board routing. (d)
Sixteen-chip cluster with an NVLink-like AIMC link and PCIe host.}
\label{fig:deployment}
\end{figure}

Fig.~\ref{fig:deployment} illustrates how a model is deployed onto a
cluster of such cards. The linear operators of each transformer block
(attention projections and FFN weight matrices) are the mapping targets:
every weight matrix is split into array-block tiles, and a remapping step
relocates weights around defective array columns using spare columns. The
resulting assignment---layer ID to (board, chip, array block)---together
with the remap table is recorded in a cluster database, which the runtime
queries to route activations across the machine. Sixteen chips are organized as SoC+AIMC boards under a top-level rack switch. The deployment diagram describes the routing of measured linear MVMs; it should not be read as evidence that the complete model executes on the dies.

\subsection{Calibrated Noise Model}
\label{sec:calibration}

\begin{figure}[!t]
\centering
\includegraphics[width=1.0\columnwidth]{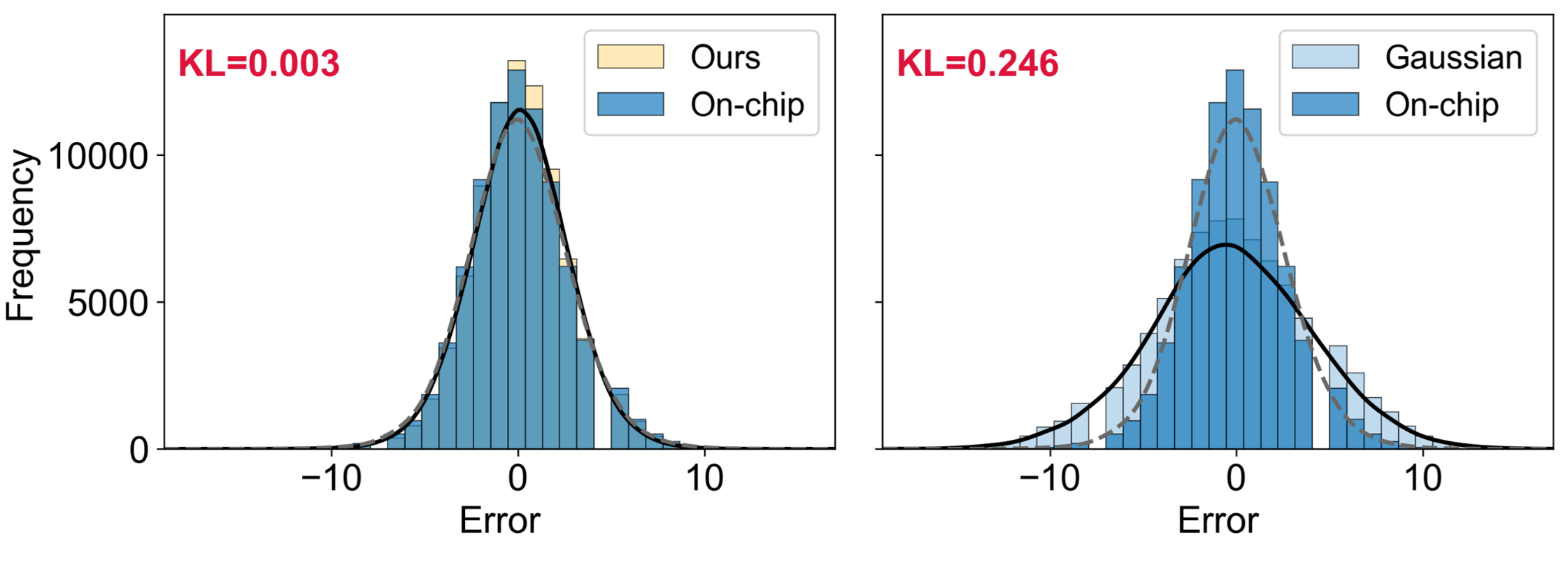}
\caption{Weight-perturbation model calibrated against measured on-chip
MVM errors (KL $=0.003$), versus an uncalibrated Gaussian (KL $=0.246$).}
\label{fig:noise_model}
\end{figure}

\begin{figure*}[!t]
\centering
\includegraphics[width=0.95\textwidth]{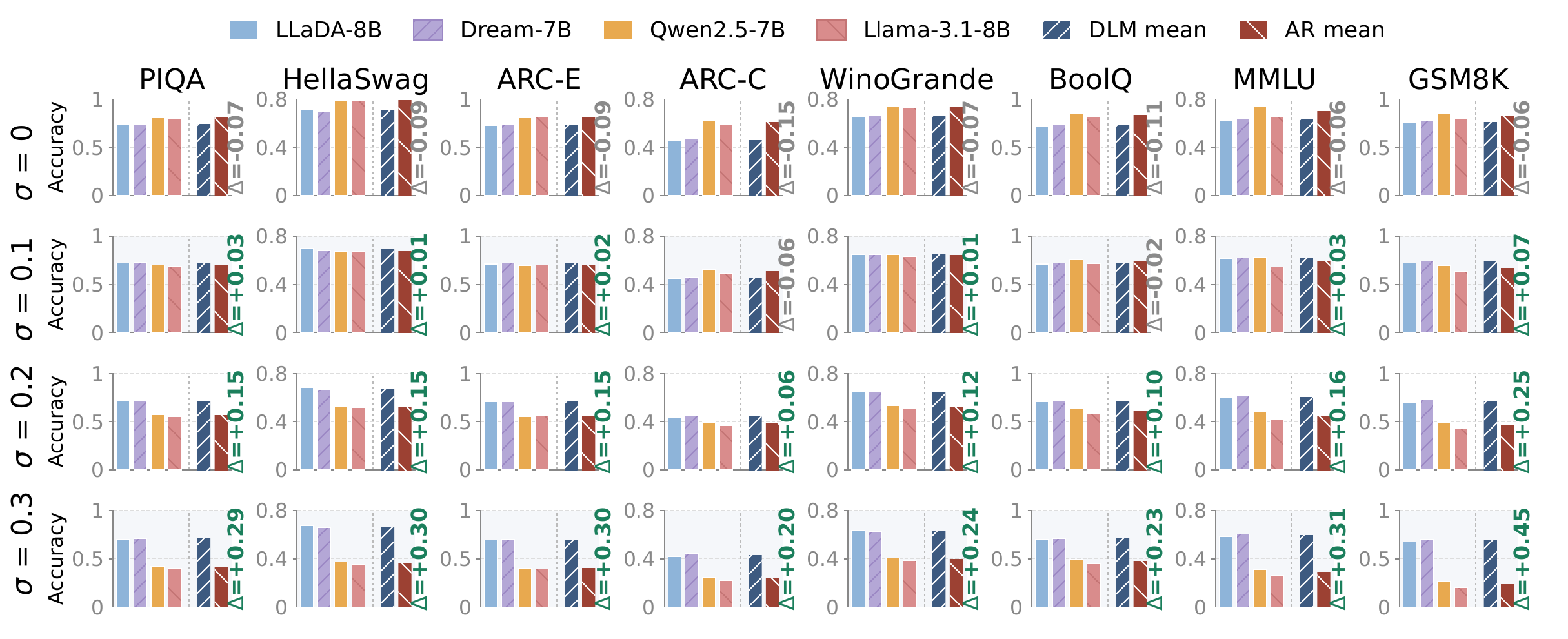}
\caption{Calibrated weight-noise sweep on an 8$\times$L20 GPU server.
Rows sweep $\sigma\in\{0,0.1,0.2,0.3\}$; bars report two DLMs (LLaDA-8B,
Dream-7B), two AR models (Qwen2.5-7B, Llama-3.1-8B), and group means.
$\Delta$: DLM-mean advantage, reaching $+0.20$--$+0.45$ at $\sigma=0.3$
on all eight benchmarks.}
\label{fig:l20_sweep}
\end{figure*}

The calibration follows NoiseGuard~\cite{DBLP:journals/tcad/WangCLLFMWZWZPTK26}, using the
weight-dependent term in Eq.~\eqref{eq:noise_distribution} with bias and TIA/ADC terms;
Sec.~\ref{sec:static_noise_eval} defines the GPU simplification and parameter
scope. The fitted marginal MVM-error histogram has KL divergence $0.003$
versus $0.246$ for an uncalibrated Gaussian
(Fig.~\ref{fig:noise_model}), an $82\times$ reduction that anchors the
calibrated perturbation model used throughout the hybrid flow. The hybrid
operating point is reported as $\sigma_{\mathrm{eff}}=0.15$.

Fig.~\ref{fig:l20_sweep} uses this calibrated model to compare DLMs and AR
models at scale in simulation on an 8$\times$L20 GPU server. Two DLMs
(LLaDA-8B, Dream-7B) and two AR models (Qwen2.5-7B, Llama-3.1-8B) are
evaluated on eight benchmarks under perturbation strengths
$\sigma\in\{0,0.1,0.2,0.3\}$. The AR models start from higher clean
accuracy at $\sigma=0$, but their advantage already vanishes at
$\sigma=0.1$ and reverses at larger perturbations: at $\sigma=0.3$ the
DLM mean exceeds the AR mean on every benchmark, with gaps ranging from
$20$ percentage points (ARC-C) to $45$ percentage points (GSM8K). This extends the observed trend to the calibrated simulation at scale.

\begin{table}[t]
\centering
\small
\setlength{\tabcolsep}{4.5pt}
\renewcommand{\arraystretch}{1.15}
\caption{Hybrid quality at $\sigma_{\mathrm{eff}}{=}0.15$ (DLMs: 256 steps).
Attention/FFN MVMs: AIMC; other computation: digital.
45M/125M: loss ($\downarrow$); 8B: GSM8K accuracy ($\uparrow$). Max $\sigma_{\mathrm{eff}}$: provisional $2\%$ relative-loss thresholds}
\label{tab:onboard}
\begin{tabular}{llcccc}
\toprule
\multirow{2}{*}{\textbf{Scale}}
& \multirow{2}{*}{\textbf{Model}}
& \multicolumn{2}{c}{\textbf{Quality}}
& \multirow{2}{*}{\textbf{$\Delta_{\mathrm{rel}}$}}
& \multirow{2}{*}{\textbf{Max $\sigma_{\mathrm{eff}}$}} \\
\cmidrule(lr){3-4}
& & \textbf{Digital} & \textbf{On-chip} & & \\
\midrule
\multirow{2}{*}{\textit{45M}}
& AR   & 3.412 & 3.451 & $+1.14\%$ & 0.19 \\
& MDLM & 3.465 & 3.494 & $+0.84\%$ & 0.27 \\
\addlinespace[3pt]
\midrule
\addlinespace[1pt]
\multirow{2}{*}{\textit{125M}}
& AR   & 3.183 & 3.207 & $+0.75\%$ & 0.28 \\
& MDLM & 3.236 & 3.242 & $+0.19\%$ & 0.58 \\
\addlinespace[3pt]
\midrule
\addlinespace[1pt]
\multirow{2}{*}{\textit{8B (open)}}
& Llama-3 & 0.748 & 0.512 & $-31.6\%$ & 0.03 \\
& LLaDA   & 0.702 & 0.628 & $-10.5\%$ & 0.09 \\
\bottomrule
\end{tabular}
\end{table}

\begin{table}[!t]
\centering
\small
\setlength{\tabcolsep}{3.2pt}
\renewcommand{\arraystretch}{1.15}
\caption{Step--quality--cost trade-off at $\sigma_{\mathrm{eff}}{=}0.15$,
with metrics and hybrid boundary as in Table~\ref{tab:onboard}.}
\label{tab:onboard_steps}
\resizebox{\columnwidth}{!}{%
\begin{tabular}{llcc>{\bfseries}ccc}
\toprule
\multirow{2}{*}{\textbf{Model}}
& \multirow{2}{*}{\textbf{Steps}}
& \multicolumn{3}{c}{\textbf{Quality}}
& \multicolumn{2}{c}{\textbf{Cost / gen.}} \\
\cmidrule(lr){3-5}
\cmidrule(lr){6-7}
& & \textbf{Digital} & \textbf{On-chip} & \textbf{$\Delta_{\mathrm{rel}}$} & \textbf{Lat.\ (s)} & \textbf{Energy (mJ)} \\
\midrule
\multirow{4}{*}{\textit{LLaDA-8B}}
& 64  & 0.381 & 0.227 & $-40.4\%$ & 0.73 & 410 \\
& 128 & 0.642 & 0.503 & $-21.7\%$ & 1.42 & 801 \\
& 256 & 0.702 & 0.628 & $-10.5\%$ & 2.81 & 1590 \\
& 512 & 0.741 & 0.712 & $-3.9\%$  & 5.58 & 3150 \\
\addlinespace[3pt]
\midrule
\addlinespace[1pt]
\multirow{4}{*}{\textit{MDLM-125M}}
& 32  & 3.251 & 3.297 & $+1.42\%$ & 0.06 & 28 \\
& 64  & 3.243 & 3.268 & $+0.77\%$ & 0.11 & 54 \\
& 128 & 3.238 & 3.250 & $+0.37\%$ & 0.22 & 107 \\
& 256 & 3.236 & 3.242 & $+0.19\%$ & 0.43 & 211 \\
\bottomrule
\end{tabular}%
}
\end{table}

\subsection{On-Board Evaluation}
\label{sec:onboard}

We evaluate at $\sigma_{\mathrm{eff}}=0.15$ with 256 DLM steps unless
specified otherwise, using the mapping flow in Fig.~\ref{fig:deployment}.
Table~\ref{tab:onboard} shows smaller degradation for MDLM at both matched
scales; at 125M, the loss increase is $0.19\%$ versus $0.75\%$ for AR.
The threshold entries give $1.4$--$3\times$ paired ratios in favor of
the DLM, widening from $1.4\times$ at 45M to $3\times$ at 8B.

Table~\ref{tab:onboard_steps} shows the DLM quality--cost trade-off:
LLaDA's relative degradation falls from $40.4\%$ to $3.9\%$ as steps
increase from 64 to 512, with approximately linear reported cost.
Standard AR decoding commits each token immediately and has no denoising-step
control; the step budget is therefore a DLM-specific deployment knob that
trades generation cost for noise resilience within the hybrid measurement
boundary.

\section{Conclusion}
\label{sec:conclusion}

We study diffusion language models under AIMC weight perturbations. DLMs degrade more gracefully than AR models due to delayed token commitment and progressive context resolution, not cross-round noise averaging. The denoising budget serves as a DLM-specific latency--robustness knob. Hybrid 16-chip AIMC cluster measurements show DLMs withstand $1.4$--$3\times$ higher weight noise, establishing diffusion generation as attractive for noise-resilient AIMC inference.

\section*{Acknowledgement}
\label{sec:acknowledgement}
This work was supported in part by the Beijing Natural Science Foundation under Grant L2607008, and in part by the Beijing Nova Program under Grant 20250484807.

\clearpage
\bibliographystyle{ieeetr}
\bibliography{bibliography}
\end{document}